# On Fractals and Fractional Calculus Motivated by Complex Systems

Awatif M. Shahin*^, Elsayed Ahmed*^, Ahmed S.Elgazzar** and Yassmin A.Omar
*Mathematics Department, Faculty of sciences Mansoura 35516, EGYPT.
**Mathematics Department, Faculty of Education, El-Arish, EGYPT
^ Author to whom requests should be sent

**Abstract:**
Complex systems (CS) are ubiquitous in nature. It is argued that fractional order (FO) calculus is more suitable to describe fractal systems. Motivated by the fractal space time theory some fractional generalizations of Scrodinger and Klein-Gordon equations are given. In many CS systems statistics is described by heavy tailed distributions e.g. fractal and Levy-Weibull ones. Here a generalized Weibull (GW) distribution is proposed to interpolate between them.

## 1. Basics of Complex Systems (CS):

**Definition (1)** [1]: A complex system consists of inhomogeneous, interacting adaptive agents.

**Definition (2):** "An emergent property of a CS is a property of the system as a whole which does not exist at the individual elements (agents) level".

Typical examples are the brain (cognition), the immune system (memory) [2,3], the economy, social systems, ecology, biological systems, earthquakes etc…

*Therefore to understand a complex system one has to study the system as a whole and not to decompose it into its constituents. This totalistic approach is the opposite of the standard reductionest one, which tries to decompose any system to its constituents and hopes that by understanding the elements one can understand the whole system.*

However it is important to notice that the two approaches (CS) and reductionest complement each other since the reductionest approach has significant successes in molecular biology and quantum systems.

CS implies that experiments should be done in vivo which is not easy so mathematical models can contribute by using mathematical models, which, hopefully, preserve the main features of the CS.

Most of the real systems are CS. Moreover they have intrinsic unpredictability which causes some "seemingly wise" decisions to have harmful side effects. Therefore we should try to understand CS to try to minimize such side effects. Here we give an example of these side effects. An interesting situation arose when some countries adopted a private sector vaccination to MMR (Measles, Mumps and Rubella) [4]. It was expected that the number of Congenital Rubella Syndrome (CRS) will decrease. However it did not and in some countries (e.g. Greece and Costs Rica) it increased. The reason can be understood as follows: This vaccination to part of the population decreases the probability of contracting the disease at young age. Consequently the probability of contracting the disease at adulthood increases. This is an example of the counterintuitive effects of some vaccination programs.

There are at least two sources for unpredictability in CS. The first is the nonlinear interactions between its agents [5] hence chaos is expected to exist. The second is that CS are open systems hence perturbation to one system may affect another related one.

Therefore CS are characterized by [6]

i) CS have emergent properties hence they should be studied as a whole.
ii) CS are open with nonlinear local interactions [7] hence:
1] Long range prediction is highly unlikely.
2] When studying a CS take into consideration the effects of its perturbation on related systems.
3] Expect side effects to any "WISE" decision.
4] Mathematical and computer models may be helpful in reducing such side effects..
iii) Optimization in CS should be multi-objective and not single objective [8].
iv) CS is very difficult to control. Targeting highly connected sites may be a useful approach [7]. Also bounded delayed distributed control is important.
v) Memory and delay effects should not be neglected in CS.
vi) Most CAS are nonlocal and decentralized hence spatial effects should be included.
viii) Randomness is an intrinsic feature in CS. Also heavy tailed distributions are frequent in CS.

**2. Fractional Equations:**

Caputo's definition for derivative of order $0 < \alpha \leq 1$ is given by

$$D^\alpha f(t) = (1/\Gamma(1-\alpha)) \int_0^t [df(s)/ds]/(t-s)^\alpha$$

Consider the following evolution equation [9]

$$df(t)/dt = -\lambda^2 \int_0^t k(t-t') f(t') dt' \quad (1)$$

If the system has no memory then $k(t-t') = \delta(t-t')$ and one gets $f(t) = f_0 \exp(-\lambda^2 t)$. If the system has an ideal memory then $k(t-t') = \{1 \text{ if } t \geq t', 0 \text{ if } t < t'\}$ hence $f \approx f_0 \cos \lambda t$. Using Laplace transform $L[f] = \int_0^\infty f(t) \exp(-st) dt$ one gets L[f]=1 if there is no memory and 1/s if there is ideal memory hence the case of non-ideal memory is expected to be given by $L[f] = 1/s^\alpha$, $0 < \alpha < 1$. In this case equation (1) becomes

$$df(t)/dt = \int_0^t (t-t')^{\alpha-1} f(t') dt' / \Gamma(\alpha) \quad (2)$$

Where $\Gamma(\alpha)$ is the Gamma function. This system has the following solution

$$f(t) = f_0 E_{\alpha+1}(-\lambda^2 t^{\alpha+1}),$$

where $E_\alpha(z)$ is the Mittag Leffler function given by

$$E_\alpha(z) = \sum_{k=0}^\infty z^k / \Gamma(\alpha k + 1)$$

It is direct to see that $E_1(z) = \exp(z), E_2(z) = \cos z$.

Following a similar procedure to study a random process with memory, one obtains the following fractional evolution equation

$$\partial^{\alpha+1} P(x,t)/\partial t^{\alpha+1} = \sum_n (-1)^n / n! \partial^n [K_n(x) P(x,t)]/\partial x^n, \qquad 0 < \alpha < 1 \quad (3)$$

where P(x,t) is a measure of the probability to find a particle at time t at position x.

We expect that (3) will be relevant to many complex adaptive systems and to systems where fractal structures are relevant since it is argued that there is a relevance between fractals and fractional differentiation [10,11].

For the case of fractional diffusion equation the results are

$$\partial^{\alpha+1} P(x,t)/\partial t^{\alpha+1} = D \partial^2 P(x,t)/\partial x^2, P(x,0) = \delta(x), \partial P(x,0)/\partial t = 0 \Rightarrow$$
$$P = (1/(2\sqrt{D t^\beta})) M(|x|/\sqrt{D t^\beta}; \beta), \quad \beta = (\alpha+1)/2 \qquad (4)$$

$$M(z; \beta) = \sum_{n=0}^{\infty} [(-1)^n z^n / \{n! \Gamma(-\beta n + 1 - \beta)\}]$$

For the case of no memory $\alpha = 0 \Rightarrow M(z; 1/2) = \exp(-z^2/4)$.

## 3. Proposed fractional equations in quantum systems:

When one studies quantum mechanics one feels a sudden change from classical mechanics and old quantum mechanics on the one hand to modern quantum mechanics (beginning with Schrodinger equation) on the other side. By this we mean that suddenly we leave the familiar concepts of position, momentum, energy, etc… and we replace them by hermitian operators which eventually gives Schrodinger equation. Can the latter equation be derived from the familiar Euler-Lagrange equations?. The answer is yes according to the scale relativity theory [12,13,14,15]. According to this theory, both space-time and scale are considered. The importance of scales is that observations and possibly physical laws depend on the scale (resolution of the experiment).

Moreover in this theory it is argued that space-time is fractal (continuous but nowhere differentiable). Accordingly Nottale has shown (among several results) that Schrodinger equation is equivalent to the familiar Euler-Lagrange equations on a fractal. Thus the gap between classical and quantum mechanics have been removed.

Scale relativity theory has two main assumptions: The first is that one should study scale ($\Delta$) in addition to space and time (x,y,z,t). The second assumption is that space time, at least in the microscale, is a fractal in the general sense i.e. it is continuous but not differentiable.

The first assumption agrees with the observation that different theories are used to describe phenomena at different scales e.g. quantum mechanics is used in the micro-scale while classical mechanics is used in the intermediate ones.

Planck's time and length play a role for scales similar to light speed in special relativity.

The second assumption implies that any displacement dX should be decomposed into left and right $dX_\pm$ which are given by

$$dX_\pm = < dX_\pm > + d\varsigma_\pm, \quad < d\varsigma_\pm > = 0, \quad < d\varsigma_\pm d\varsigma_\pm > \propto (dt)^{2/D}$$

where D is the space time fractal dimension.

Applying Ito calculus procedure, it has been shown [12,13] that Schrodinger equation is equivalent to Euler-Lagrange equation on a fractal.

There is a relation between chaos and the non-differentiability of space time since if a straight line x=0,y=0,z=at, a constant is perturbed to
$x = \varepsilon_1(1+\exp(t/\lambda))$, $y = x = \varepsilon_2(1+\exp(t/\lambda))$, $z = at + x = \varepsilon_3(1+\exp(t/\lambda))$
where $\lambda$ is Lyapunov exponent of the system then it is direct to see (by eliminating the time) that dz/dx is not defined as x→0. This may be related to the existence, in many phenomena, of a horizon of predictability.

Nottale [12,13] has applied the Schrodinger type equation resulting from scale relativity to gravity and obtained
$$D^2\nabla^2\psi + iD\partial\psi/\partial t + GM\psi/r = 0$$
$|\psi|^2$ is a measure for making structures e.g. planets, stars, galaxies, etc… and used it to describe some astrophysical phenomena.

We think that the following points makes scale relativity interesting and deserves further research:
1) The dependence on scale is an observed phenomena.
2) It relates classical and quantum mechanics in a natural way by proving that Schrodinger equation is equivalent to Euler -Lagrange equation on a fractal.
3) Since earthquakes mostly occur on fractal faults, applying scale relativity to the earthquake phenomena is an interesting application [Ahmed and Mousa 1997].
4) Since it has been found that fractional calculus is more natural to apply on fractals [10,11] generalizing scale relativity equations to fractional order should be done. It is interesting to know that several attempts have been done to generalize Schrodinger equation to fractional order as will be seen next section..

According to scale relativity theory, fractals are expected to be a relevant concept in quantum systems. It is known that the fractional equations (FE) are the natural tools to study fractal systems. Therefore it is relevant to attempt to use FE to generalize both Schrodinger and Klein-Gordon equations.

In the next section it will be shown that FE are also relevant to systems with memory which typically are open and dissipative systems and to complex systems [16].

Motivated by the previous two sections we propose the following fractional equation to describe quantum systems:
$$e^{i\pi\beta}h\partial^{2\beta}\Psi/\partial t^{2\beta} = (h^2/2m)(-\Delta)^\gamma\Psi + V\Psi \quad (5)$$
where
$\beta = (1+\alpha)/2$, $\alpha$ is in (4), h is related to Planck constant, $\Psi$ is the wave function of the system V is the potential and $\Delta$ is Laplace operator.

It is clear that (5) reduces to the familiar Schrodinger equation if ($\alpha \to 0$, $\gamma \to 1$) and is equivalent to Klein-Gordon equation as ($\alpha \to 1$, $\gamma \to 1$).

Other generalizations have been proposed by Chen [17] where both space, time and even potential are fractionalized
$$e^{i\pi\beta}h\partial^{2\beta}\Psi/\partial t^{2\beta} = (h^2/2m)(-\Delta)^{2\beta}\Psi + V\Psi$$
$$e^{i\pi\beta}h\partial^{2\beta}\Psi/\partial t^{2\beta} = (h^2/2m)(-\Delta)^{2\beta}\Psi + V^{2\beta}\Psi \quad (6)$$

Recently [18] the fractional potential well has been solved. It is given by (up to constants) (5) with potential function $V(x) = \{0 \text{ if } 0 < x < a, \infty \text{ otherwise}\}$

Thus using separation of variables one gets $\Psi = f(t)\sin(n\pi x/a)$, n = 1,2,3,.... It is straightforward to see that $f(t) = E_{2\beta}(\omega_n(-it)^{2\beta})$ where $E_{2\beta}$ is Mittag-Leffler function and $\omega_n$ are parameters. Using Naber's formula

$$E_\nu(\omega(-it)^\nu) = (1/\nu)\{\exp(-i\omega^{1/\nu}t) - \nu F_\nu((-i\omega)^\nu, t)\} \quad (7)$$

where F decays monotonically with time. Thus as time increases the first term in (7) becomes dominant. Consequently the energy of the nth level is proportional to $\varepsilon_n \propto n^{2\beta}$. Similar argument leads us to expect that for the fractional hydrogen atom the energy of the nth level is
$\varepsilon_n \propto (-1/n^{2\beta})$ and that the radiation emitted or absorbed in transition between energy levels is proportional to $((1/n^{2\beta}) - (1/k^{2\beta}))$ where k > n are positive integers.
Thus the fractional approach affects the energy levels but it preserves quantization.

Another application of fractal space time is to model earthquakes [19,20] since it is known that earthquakes occur on faults which are mostly fractals. Generalizing to fractional models for earthquakes has already been done [20].

Recently [13] a quantum space time theory has been proposed and was successful in deriving mass spectrum of QED and QCD particles. It also depends on the assumption that at micro scale, space time has a hierarchical Cantor like fractal structure. Therefore we expect FE to be relevant to this theory.

It is also expected that the intrinsic non locality of fractional calculus may explain the entanglement problem in quantum mechanics.

**4. Generalized Weibull distribution (GW):**

As has been stated in sec.1, heavy tailed distributions are frequent in CS [6]. There are two extremes of such distributions namely fractal distributions given by the probability density function:
$$f(x) = cx^{-\alpha}, \alpha > 1$$
and the Levy-Weibull one with probability density function:
$$f(x) = cx^{\beta-1}\exp(-x^\beta), \beta > 0$$
Here we propose a generalized Weibull distribution (GW) with probability density function
$$f(x) = cx^\gamma \exp(-x^\beta), \quad (8)$$
where $\beta \geq 0, \gamma$ can be positive or negative. It is clear that it includes both distributions. Hence it can be used to describe fractal systems. Also it includes the cases where there is a finite number of random variables with extreme values (e.g. earthquakes). In this case the distribution is given by [21,22] Levy-Weibull distribution.

Now are there phenomena represented by the proposed GW distribution? The answer is yes e.g. scaling in percolation theory [23]. Another example is an extinction model [Ahmed et al 2000] which has $c \approx 22.17, \gamma \approx 1.37, \beta \approx .76$.

Summarizing: Complex systems (CS) are frequent in nature. Mathematical methods can be helpful in understanding them. Motivated by the properties of CS we concluded that:
i) Fractional order systems are more suitable for describing CS for the following reasons: First they are more natural in describing fractal systems. Second they are more natural in describing systems with memory and delay. Third they are more natural in describing non-local systems.

      This may have an impact on the entanglement problem in quantum mechanics.
- ii) Motivated by the theory of fractal space time, fractional order generalization for Schrodinger equations are presented.
- iii) Finally a generalized Weibull distribution is proposed as a generic distribution for several complex systems.

It is interesting that the bright idea of Nottale, Ord and El-Naschie namely fractal space time offers a unifying scheme for such diverse fields as quantum mechanics, earthquakes and fractional order equations

## 6. Conclusions:

Complex systems (CS) are frequent in nature. Mathematical methods can be helpful in understanding them. Motivated by the properties of CS we concluded that:
- i) Fractional order systems are more suitable for describing CS for the following reasons: First they are more natural in describing fractal systems. Second they are more natural in describing systems with memory and delay. Third they are more natural in describing non-local systems. This may have an impact on the entanglement problem in quantum mechanics.
- ii) Motivated by the theory of fractal space time, fractional order generalization for Schrodinger equations are presented.
- iii) Finally a generalized Weibull distribution is proposed as a generic distribution for several complex systems.